\documentclass[12pt]{article}

\usepackage{graphics}

\title{Observation of Single Transits in Supercooled Monatomic Liquids}
\author{Duane C.\ Wallace, Eric D.\ Chisolm, and Brad E.\ Clements \\
        Theoretical Division \\ Los Alamos National Laboratory \\Los 
        Alamos, NM~~87545}

\begin{document}

\maketitle

\begin{abstract}
A {\em transit} is the motion of a system from one many-particle
potential energy valley to another.  We report the observation of
transits in molecular dynamics (MD) calculations of supercooled liquid
argon and sodium.  Each transit is a correlated simultaneous shift in
the equilibrium positions of a small local group of particles, as
revealed in the fluctuating graphs of the particle coordinates versus
time.  This is the first reported direct observation of transit motion
in a monatomic liquid in thermal equilibrium.  We found transits
involving 2 to 11 particles, having mean shift in equilibrium position
on the order of $0.4\,R_1$ in argon and $0.25\,R_1$ in sodium, where
$R_1$ is the nearest neighbor distance.  The time it takes for a
transit to occur is approximately one mean vibrational period,
confirming that transits are fast.
\end{abstract}

\section{Introduction}
\label{intro}

Long ago, Frenkel \cite{1,2} noted that the liquid-solid phase
transition has only a small effect on volume, cohesive forces, and
specific heat, while the liquid \mbox{diffuses} much more rapidly than
the solid, and from these facts he argued that the motion of a liquid
atom consists of approximately harmonic oscillations about an
equilibrium point, while the equilibrium point itself jumps from time
to time.  Goldstein \cite{3} pictured the motion of liquid atoms as
primarily controlled by thermal activation over barriers, with a
distribution of barrier heights.  From computer simulations,
Stillinger and Weber \cite{4,5} found mechanically stable
arrangements of particles, called inherent structures, and they
suggested that the equilibrium properties of liquids result from
vibrational excitations within, and shifting equilibria between, these
inherent structures.  Their simulations showed a range of energies for
these structures, so when Stillinger and Weber formulated a
statistical mechanics of their system they included a distribution of
inherent structure potential energies \cite{6,7}.  Since then, the
picture has developed of a ``rugged potential energy landscape,'' with
a wide distribution of structural potential energies, separated by
barriers having a wide distribution of heights \cite{8,9,10,11}.  Here
we use the term ``structure'' to indicate any mechanically stable
configuration of particles, corresponding to a local minimum in the
many-particle potential surface.

The present study is limited to monatomic liquids, meaning elemental
liquids which do not have molecular bonding.  Monatomic liquids
include all elemental liquid metals and the rare gas liquids, but not
the molecular liquids N$_2$, O$_2$, etc., and not polyatomic systems
such as alkali halides or water.  Molecular liquids have
translational, rotational, and vibrational degrees of freedom, while
monatomic liquids have only translational motion, and the potential
energy surface for monatomic liquids is presumably the simplest of all
liquid potential landscapes.  We use the word ``ion'' as in metals
theory, where an ion consists of a nucleus plus a rigid electron core.

The present database of thermodynamic properties of crystals and
liquids, much more extensive and accurate than was available to
Frenkel, suggests a potential energy surface for monatomic liquids
much simpler than the rugged landscape picture above.  Two crucial
pieces of experimental information, summarized below, lead to this
conclusion.  First, the constant volume specific heat due only to the
motion of the ions, denoted $C_I$, is approximately $3 k_B$ per ion.
This is true for every liquid metal for which the necessary data are
available (see Table I of \cite{12} or Fig.\ 1 of \cite{13}), and
there is no experimental indication that this property fails for any
liquid metal in the periodic table.  In addition, although liquid Ar
at 1 bar is somewhat gaslike \cite{12}, $C_I$ is close to $3 k_B$
for compressed liquid Ar \cite{13}, so it is included in our list of
monatomic liquids.  The property $C_I \approx 3 k_B$ strongly suggests
that the ions spend most of their time moving within nearly harmonic
many-particle potential energy valleys.

The second piece of evidence is the entropy of melting, but before
this information can be made quantitative, one must recognize two
categories of melting of elements \cite{101}, namely (a) normal
melting, in which the electronic structure of crystal and liquid are
the same (e.g.\ metal to metal) and (b) anomalous melting, in which
the electronic structure changes significantly upon melting (e.g.\
semiconductor to metal).  Then the constant volume entropy of melting
for the normal melting elements is found to be a universal constant,
and again this property holds without exception for all the elements
for which sufficient experimental data exist, including compressed
liquid Ar \cite{101,102,103}.  Unlike the specific heat data, the
entropy of melting does not compel us to an immediate conclusion, but
we can construct an interpretation consistent with the data.  The
interpretation proposed in \cite{12} is that the potential valleys
important in the statistical mechanics of monatomic liquids are all
alike, with each having the same structural potential $\Phi_0$ and
distribution $g(\omega)$ of harmonic normal mode frequencies.  On the
other hand, we know from Stillinger and Weber \cite{6,7} that a
distribution of $\Phi_0$ values can be seen in computer simulations,
and we also know that crystalline valleys of different symmetry have
different $\Phi_0$ values, so in \cite{12} we conjecture that the
potential valleys fall into two classes, namely (a) symmetric valleys,
which have some crystalline short-range order and hence have a
distribution of $\Phi_0$ values, and (b) random valleys, which have no
order parameter and hence all have the same shape in the thermodynamic
limit (same $\Phi_0$ and $g(\omega)$), and which are of overwhelming
numerical superiority relative to the symmetric valleys.  Then the
statistical mechanics of the liquid state depends only on the random
valleys, and the universal entropy of melting is simply related to a
very large universal number of random valleys.  From this description
of the potential surface, the Hamiltonian can be written and the
partition function evaluated, and an accurate account of thermodynamic
properties of monatomic liquids is obtained \cite{12,13}.  More
recently \cite{14,15}, computer simulations of sodium have provided a
detailed verification of this description of the many-particle
potential energy surface.  (That all random valleys have the same
$\Phi_0$ is shown by Eq.\ (3.3) of \cite{14}, and a demonstration that
all random valleys have the same $g(\omega)$ is found in Fig.\ 7, and
in the discussion surrounding Eq.\ (3.7), of \cite{14}.)  A similar
verification, but less detailed, has been obtained for Lennard-Jones
argon \cite{16}.

This description of the potential surface has two important
implications for the motion of the system, called a {\em transit},
when it passes from one many-particle valley to another.  First,
because of the role of transits in establishing and maintaining
equilibrium, transits must be local; i.e.\ each transit must involve
only a small localized group of particles \cite{12} (except for coherent
quantum states, not under consideration here).  Second, because the
transit motion has little effect on the ion motional specific heat,
transits must be sharp, i.e.\ of short time duration.  A model of
instantaneous transits has been applied to the velocity
autocorrelation function and self diffusion \cite{17,105}, and the idea
that transits are correlation controlled, as opposed to thermally
activated, has been applied to the glass transition \cite{18}.  The
purpose of the present paper is to report the observation of
individual transits in molecular dynamics (MD) calculations for
monatomic systems of argon and sodium.  Our procedure and results are
given in Sec.\ \ref{obstrans}, and a comparison with previous results
is given in Sec.\ \ref{discuss}.

This work is the first observation of individual transits as they
appear in the actual trajectory of an equilibrium MD system when it
passes from one many-particle valley to another.  A different
technique, called ``inherent dynamics,'' which maps successive
configurations of an MD calculation onto a time series of inherent
structures \cite{6,7,eh}, has yielded results in some ways complementary
to the present study.  Inherent dynamics was applied to a binary
Lennard-Jones system by Schr{\o}der et al.\ \cite{104} to show that as
temperature decreases toward a crossover temperature $T_x$ the self
part of the intermediate scattering function decays at two distinct
relaxation times, a short vibrational relaxation time and a long
relaxation time associated with transitions between inherent
structures.  Schr{\o}der and coworkers located transitions between
inherent structures by monitoring the inherent structure potential,
our $\Phi_0$, and also the real-space location of the inherent structure as
functions of time.  This observation demonstrates that the MD system,
moving in equilibrium, will quench into different inherent structures
at different times, but it does not tell us about the actual system
motion during a transit.  The present study is intended to provide
insight into that process.

\section{Observation of Transits}
\label{obstrans}

We searched for transits in 500 particle systems with periodic
boundary conditions.  To reveal the most detailed and precise
expression of the transit process, we monitored all Cartesian
coordinates of all particles as functions of time during an
equilibrium MD run.  As mentioned in \cite{14}, at a sufficiently low
temperature, the system moves within a single random valley for as
long as we can continue the MD run.  In this event, the graph of each
coordinate of each particle is a fluctuating signal with constant
mean, where the mean value locates the particle's equilibrium
position, and the set of all such graphs constitutes an unambiguous
observation that the system is moving in a single potential valley.
In the present work, the temperature was chosen so that transits occur,
but rarely, so that the graph of each coordinate of each particle is
again a fluctuating signal with constant mean for some time, then a
shift appears in the mean coordinates of several particles, and then
the graphs continue as fluctuating signals with constant means.  These
graphs constitute an unambiguous observation that the system moves for
a time within a single potential valley, then transits to a new
valley, then continues to move within the new valley.  Having thus
isolated transits in the equilibrium MD motion, we can study their
properties, such as how much time they take, how many particles are
involved, and how far their equilibrium positions shift.  Certain
characteristics of our study should be mentioned at the outset.
First, throughout each equilibrium run, those with transits and those
without, the mean potential and kinetic energies of the system showed
no perceptible change, hence every transit observed is between two
random valleys.  (Recall from Sec.\ \ref{intro} that all random
valleys have the same depth $\Phi_0$.)  Second, the mean Cartesian
coordinates of every particle were constant throughout each
equilibrium run, except for transits.  In other words, no motion other
than equilibrium vibrations and transits occurred.  Finally, the
graphs shown are representative of all the graphs we observed, and no
selection of ``best examples'' was necessary. 

During each equilibrium run, we identified a potential transit when
the running average of any coordinate over the 5000 previous timesteps
moved by a distance equal to or greater than a prescribed criterion
(listed below).  Upon inspection, we then verified that in every case
we identified, the coordinates of more than one particle moved at the
same time in the manner described above, indicating a genuine transit,
and every transit was from one random valley to another.

The density of our Lennard-Jones system is $0.9522$
particles/$\sigma^3$, with corresponding nearest neighbor distance
$R_1=1.095\,\sigma$, taken as the first maximum of $g(r)$ in the
liquid state.  When applied to argon ($\sigma = 3.405$ \AA), the
density is $1.600$ g/cm$^3$, the rms normal mode frequency of the
random valleys is $6.88 \times 10^{12}$ s$^{-1}$, and the mean
vibrational period is $\tau = 424\,\delta t$, where the MD timestep is
$\delta t = 2.15634$ fs.  (For comparison, the density of liquid Ar at
1 bar is 1.414 g/cm$^3$.)  The forces and potentials in the system are
computed taking into account all pairs of particles, using the full
Lennard-Jones potential.  The transit criterion is $0.1\,\sigma$, or
approximately $0.1\,R_1$ (that is, all motions greater than $0.1\,R_1$
in any coordinate averaged over time were tagged as potential
transits), and the lowest temperature where transits were observed was
$17.1$ K, roughly comparable to the glass transition temperature.

\begin{figure}[p]
\includegraphics{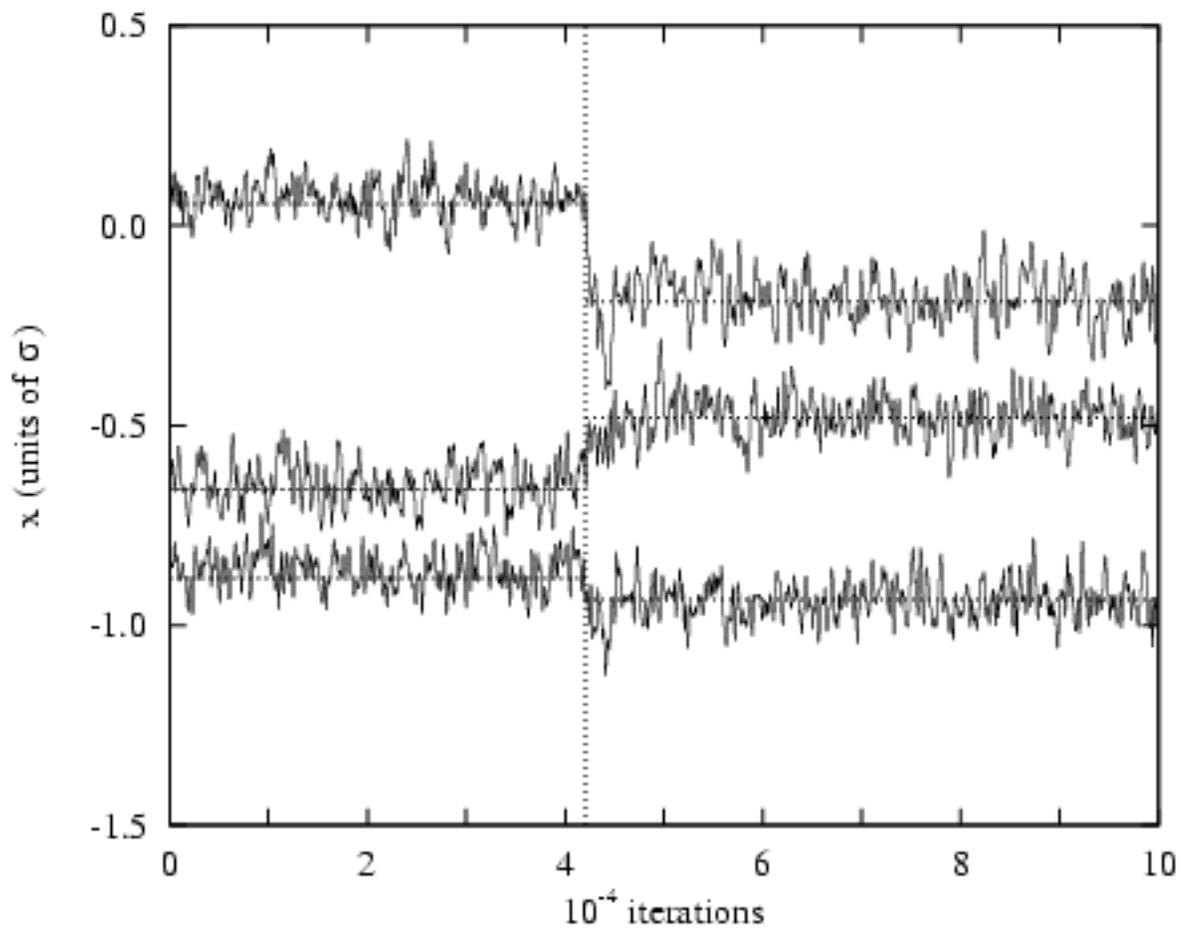}
\caption{The $x$ coordinates of (top to bottom) the seventh, fifth, and second 
         particles involved in an $8$-particle transit in Lennard-Jones argon 
         at $17.1$ K.  The transit time is the same as in Figs.\ \ref{fig2} and
         \ref{fig3}.}
\label{fig1}
\end{figure}

\begin{figure}[p]
\includegraphics{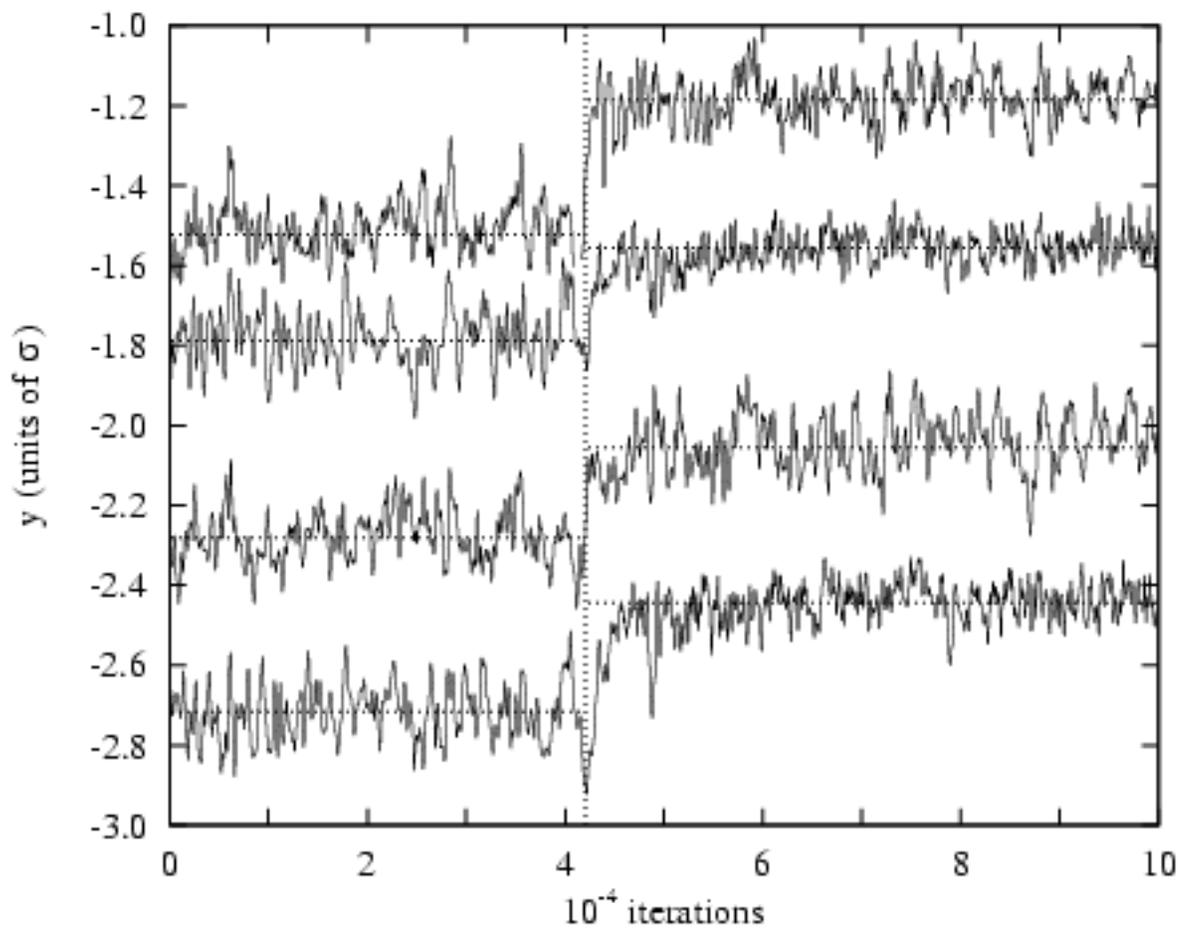}
\caption{The $y$ coordinates of (top to bottom) the first, second, eighth, and 
         sixth particles involved in an $8$-particle transit in Lennard-Jones 
         argon at $17.1$ K.  The transit time is the same as in Figs.\ 
         \ref{fig1} and \ref{fig3}.}
\label{fig2}
\end{figure}

\begin{figure}[p]
\includegraphics{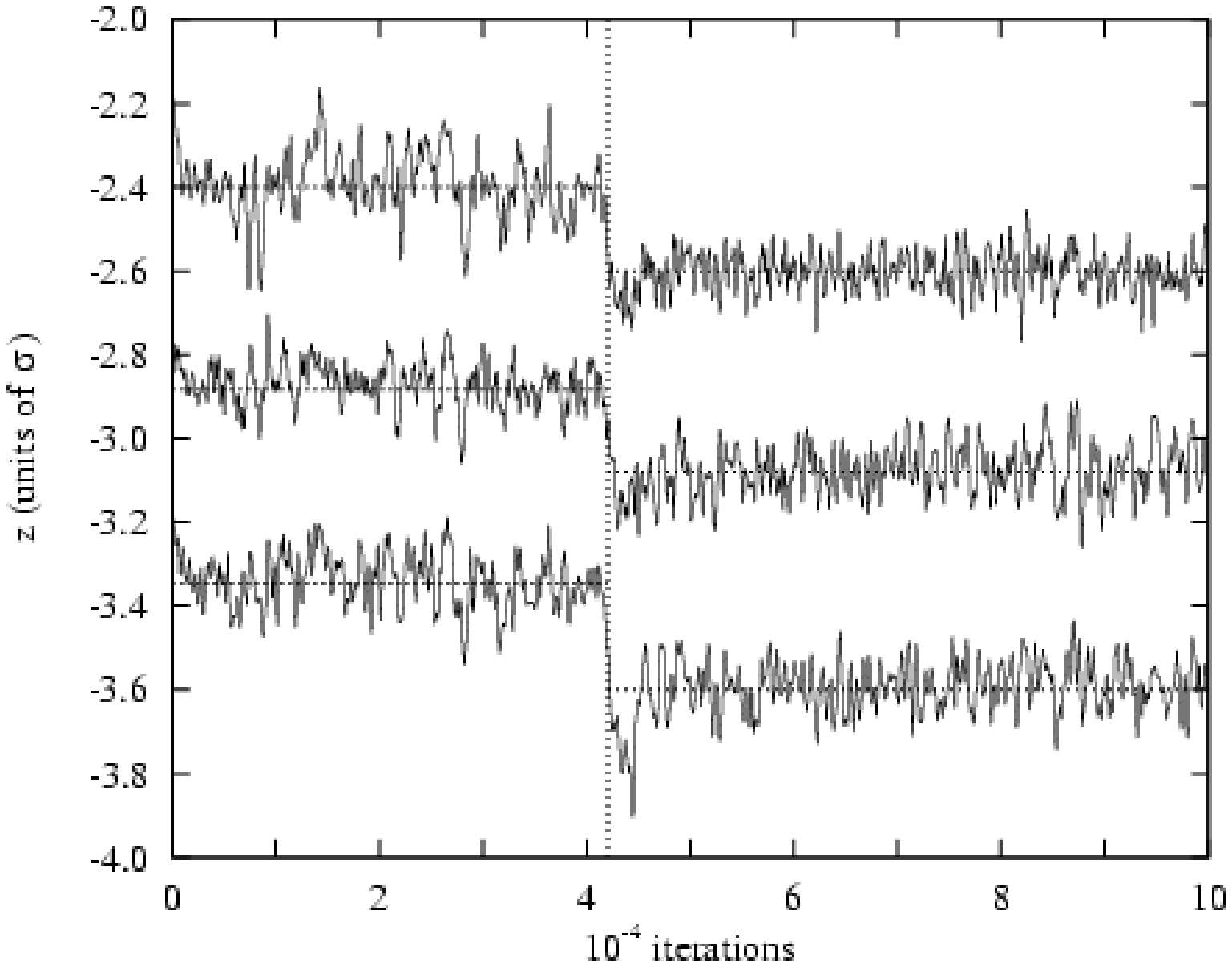}
\caption{The $z$ coordinates of (top to bottom) the first, seventh, and second 
         particles involved in an $8$-particle transit in Lennard-Jones argon 
         at $17.1$ K.  The transit time is the same as in Figs.\ \ref{fig1} and
         \ref{fig2}.}
\label{fig3}
\end{figure}

Figs.\ \ref{fig1}, \ref{fig2}, and \ref{fig3} show respectively some of 
the $x$, $y$, and $z$ coordinates of an 8-particle transit in argon at 
$17.1$ K.  
Only some particles are shown in each Figure for the sake of clarity,
and different particles are shown in different Figures, but those that
are shown are representative (the particle numbers are listed in the
Figure captions).  While Figs.\ \ref{fig2} and \ref{fig3} suggest that
all transiting particles move in the same direction, this is not
actually the case, and the appearance results from keeping only a set
of clearly distinguishable curves.  (When all the curves are plotted
together, the motion of individual particles is difficult to see.)
The dotted lines are drawn for visual guidance.  The vertical line
indicates the transit time, which is the same in all three Figures.
The system is undergoing harmonic vibrational motion in one random
valley before the transit, and in another random valley after the
transit.  The displacement of a Cartesian component of the equilibrium
position of a particle is given by the change in the horizontal dotted
line.  To estimate the duration of a transit for each particle
separately, we draw horizontal lines approximating the upper and lower
bounds of each fluctuating signal, and find how long the transiting
particle is outside of both its pre-transit and post-transit bounds.
By this measure, many of the graphs in Figs.\ \ref{fig1}-\ref{fig3}
show zero transit duration.  A close examination reveals that the best
choice for the transit time, as well as the transit duration, varies
slightly even among the three coordinates of one particle.  Our
practice is to set the transit time precisely the same for all
coordinates of all particles, and allow the transit duration to cover
remaining variations.  Let us denote by $\Delta R$ the distance over
which the equilibrium position of a transiting particle moves, and by
$\Delta t$ its transit duration.  Then for the $8$ particles involved
in the transit, $\Delta R$ varies from $0.3\,R_1$ to $0.6\,R_1$, with
a mean value of $0.4\,R_1$, and $\Delta t$ has an estimated mean value
of $\tau$.

\begin{figure}[p]
\includegraphics{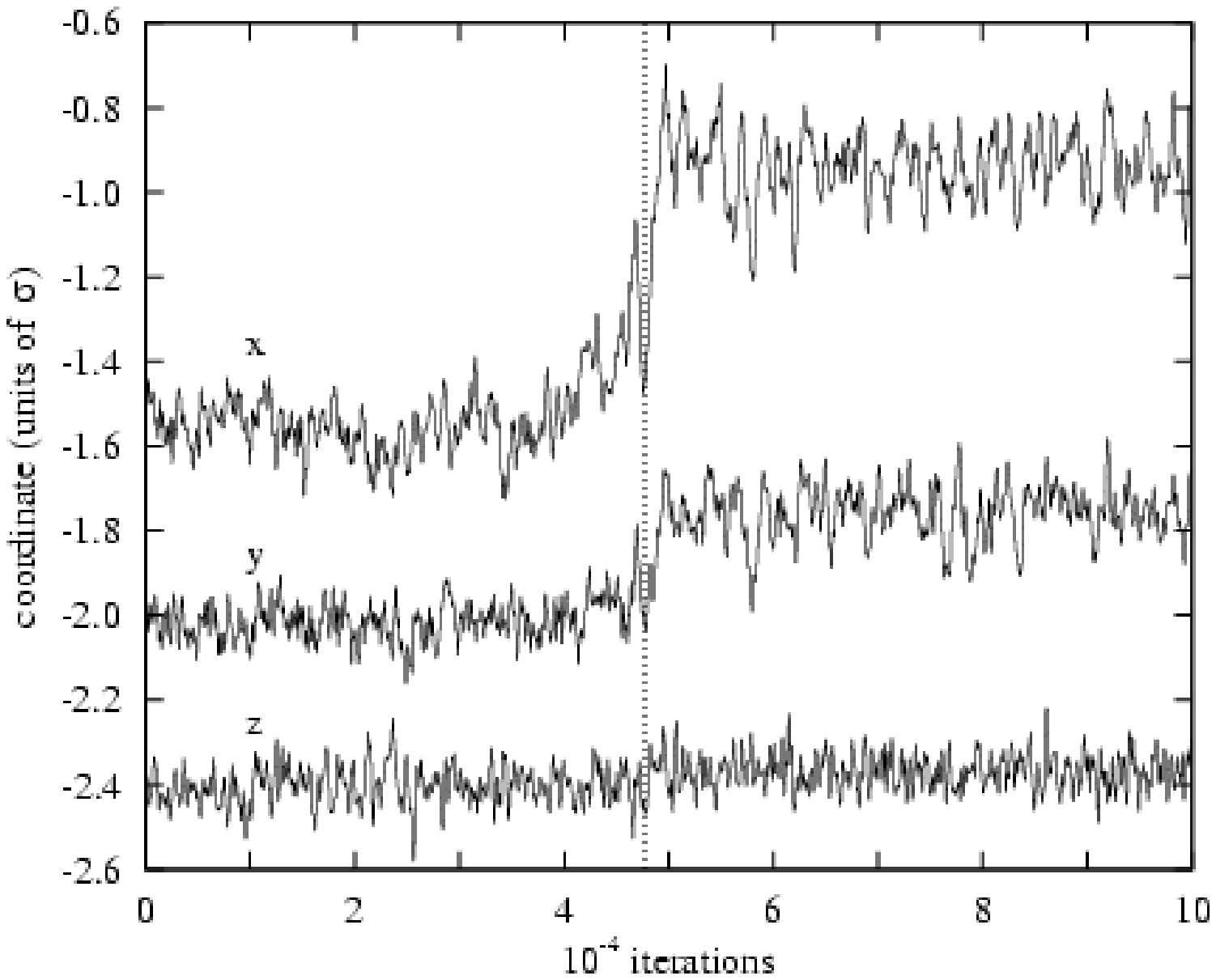}
\caption{The coordinates of one of three particles involved in a later transit 
         in Lennard-Jones argon at $17.1$ K.  The $y$ and $z$ coordinates have 
         been shifted for clarity.}
\label{fig4}
\end{figure}

Following the $8$-particle transit by a time of $13\,\tau$, another
transit occurred among three entirely different particles.  The
coordinates of one of these are shown in Fig.\ \ref{fig4}.
There is a small but measurable change in the mean of $z$, a nominal
change in the mean of $y$, and a large change in the mean of $x$.  At
the transit there is a slight decrease in the $z$ vibrational
amplitude, and a noticeable increase in both the $x$ and $y$
amplitudes.  Such changes in amplitude are common in transits we
observed, but of course they must average away over many transits,
because these amplitudes are all selected from a single equilibrium
distribution.  (Notice Figs.\ \ref{fig1}-\ref{fig3} also exhibit a
distribution of amplitudes.)  The $x$ coordinate in Fig.\ \ref{fig4}
shows a significant precursor, unusual but not singular, extending
ahead of the transit time, while the $y$ coordinate shows only the
hint of a precursor.

\begin{figure}[p]
\includegraphics{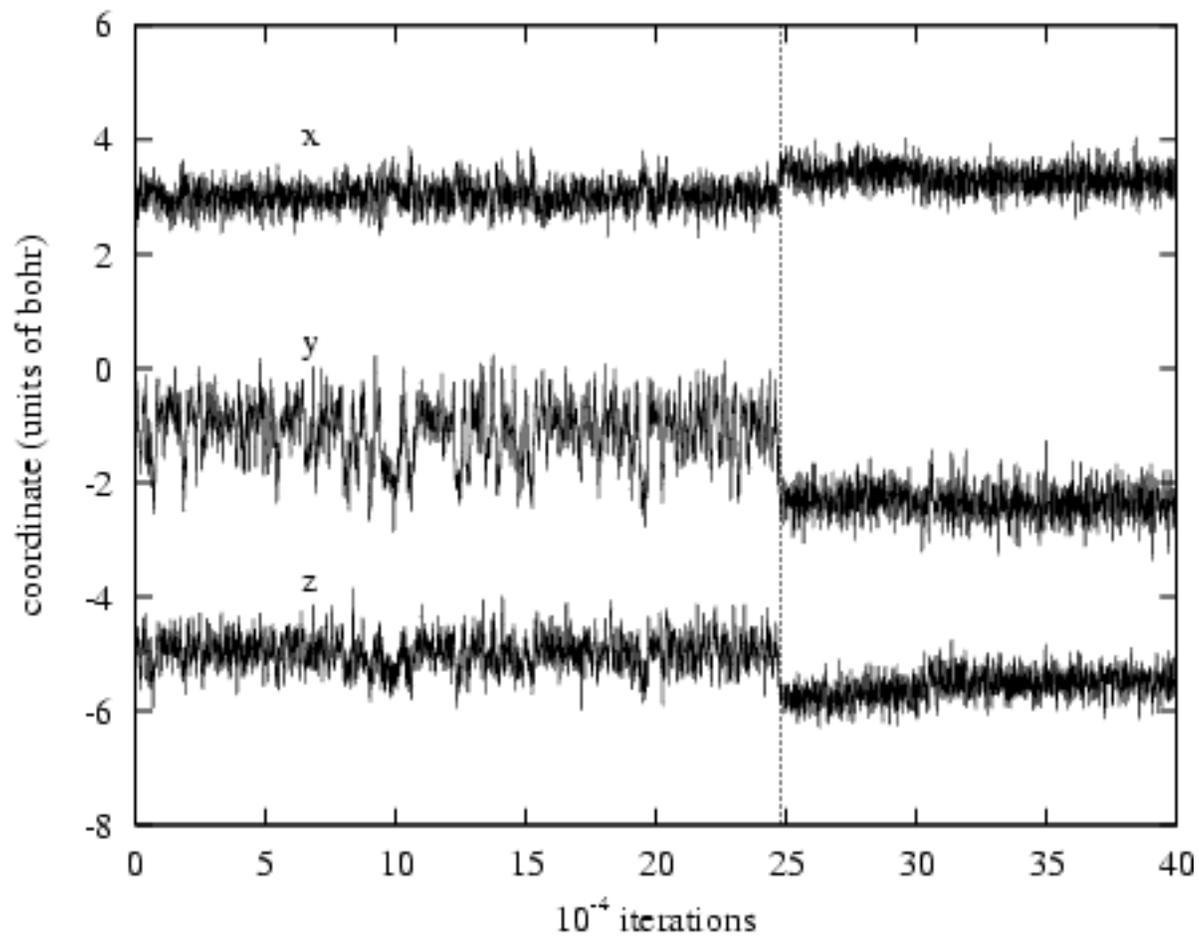}
\caption{The coordinates of one particle in an $11$-particle transit in sodium 
         at $30.0$ K.}
\label{fig5}
\end{figure}

Our sodium system has potential energy based on pseudopotential
theory, and is described in \cite{14,15}.  The density corresponds to
liquid sodium at $T_m=371$ K, and the nearest neighbor distance is
$R_1=7.0$ bohr.  The rms vibrational frequency of the random valleys
is $1.562 \times 10^{13}$ s$^{-1}$, and the mean vibrational period is
$\tau = 287.25\,\delta t$, where the MD timestep is $\delta t
=1.40058$ fs.  The transit criterion is $1$ bohr, or $0.14\,R_1$, and
the lowest temperature where transits were observed was $30.0$ K, roughly
$30\%$ of the glass transition temperature.  At this temperature we
observed a transit involving $11$ particles, and the set of graphs of
particle coordinates versus time is qualitatively indistinguishable
from the argon graphs shown in Figs.\ \ref{fig1}-\ref{fig3}.  The
three coordinates of one of the transiting particles are shown in Fig.\
\ref{fig5}, and the remaining $10$ particles exhibit similar and equally
striking graphs.
Again for this $11$-particle transit in sodium, every coordinate of
every particle transits at the same time.  The mean single-particle
transit distance is $0.25\,R_1$ while the group center of mass moves a
distance $0.11\,R_1$, and the mean single-particle transit duration is
approximately $\tau$.  The $11$ transiting particles occupy a roughly
cubical volume, which contains altogether $23$ particles, so that the
transiting group has a rather compact shape.
	
Recall that in identifying each transit, we find every particle for
which at least one running average coordinate moves as much as the
transit criterion.  Upon reducing the transit criterion by half in
sodium, to $0.07\,R_1$, we found many more particles participating in
each transit, but we found no new transits.  This suggests there are
no transits which have only very small positional shifts $\Delta R$,
but when a transit does occur, many surrounding particles undergo
small correlated positional shifts.  In Fig.\ \ref{fig5}, the small
shifts in $x$ and $z$ at approximately $310,000$ iterations are
associated with the transit of another group of particles.

\begin{figure}[p]
\includegraphics{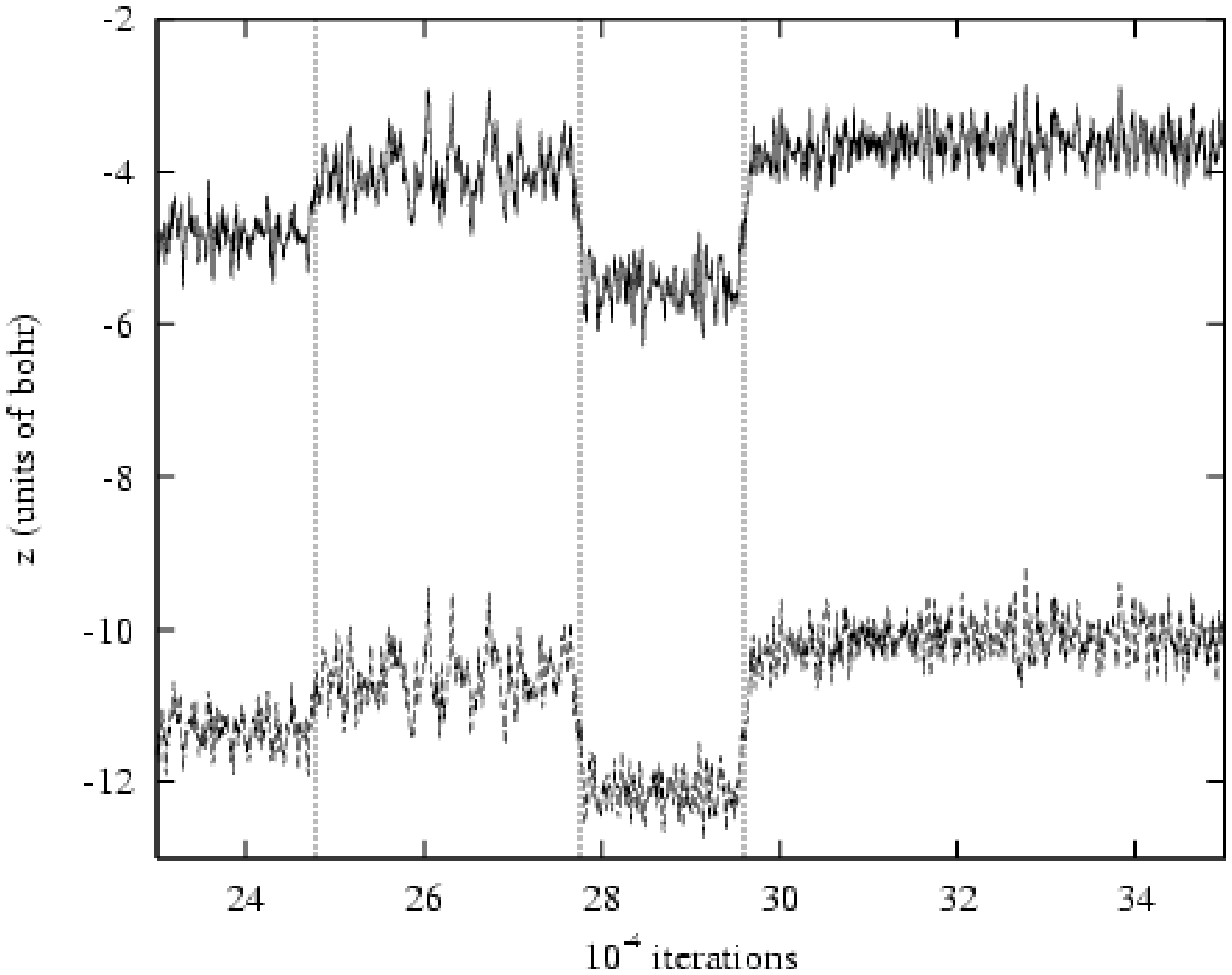}
\caption{The $z$ coordinates of two sodium particles involved in three separate
         transits over a period of $120,000$ iterations.  The transit times are
         the same as in Figs.\ \ref{fig7} and \ref{fig8}.}
\label{fig6}
\end{figure}

\begin{figure}[p]
\includegraphics{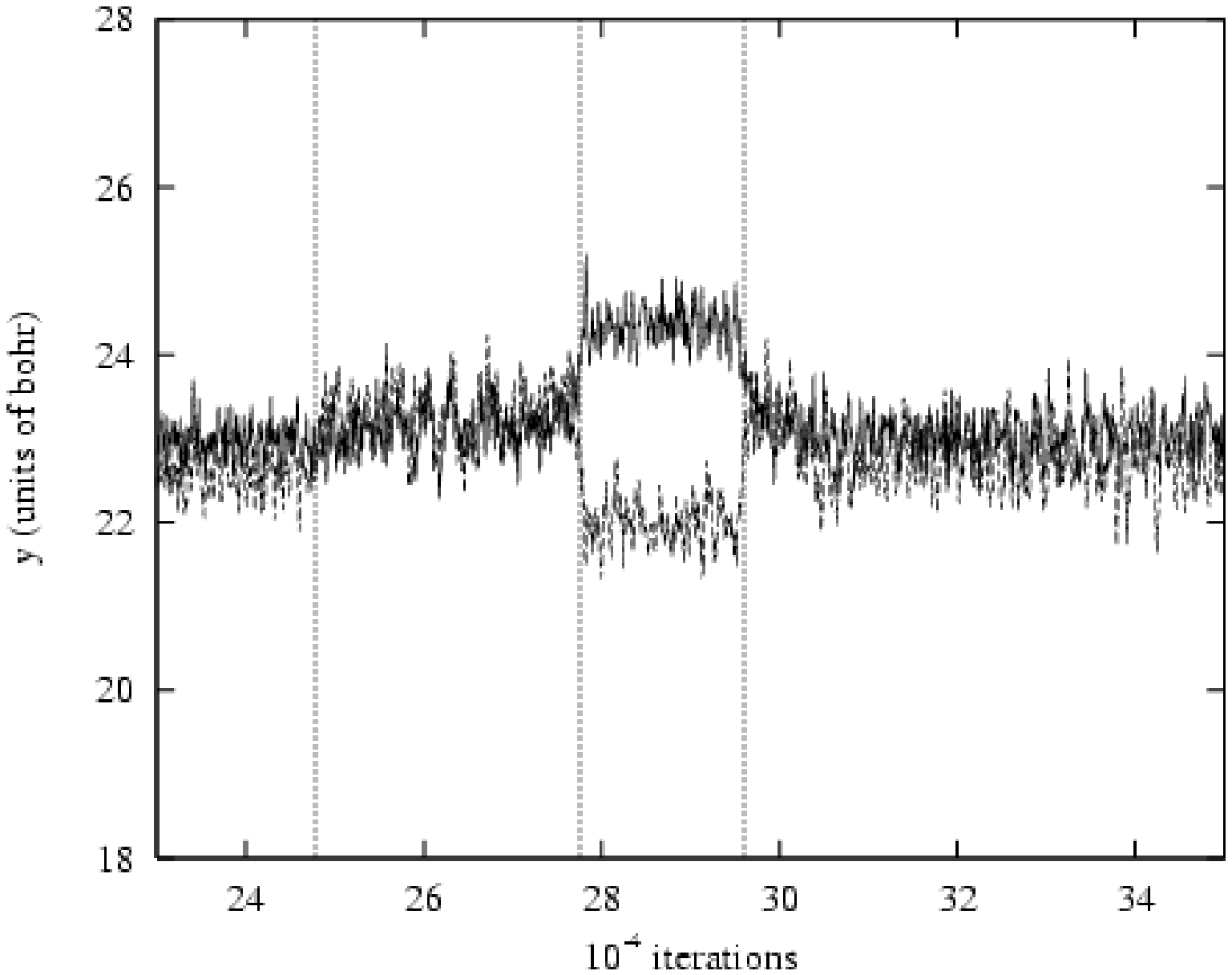}
\caption{The $y$ coordinates of two sodium particles involved in three separate
         transits over a period of $120,000$ iterations.  The transit times are
         the same as in Figs.\ \ref{fig6} and \ref{fig8}.}
\label{fig7}
\end{figure}

\begin{figure}[p]
\includegraphics{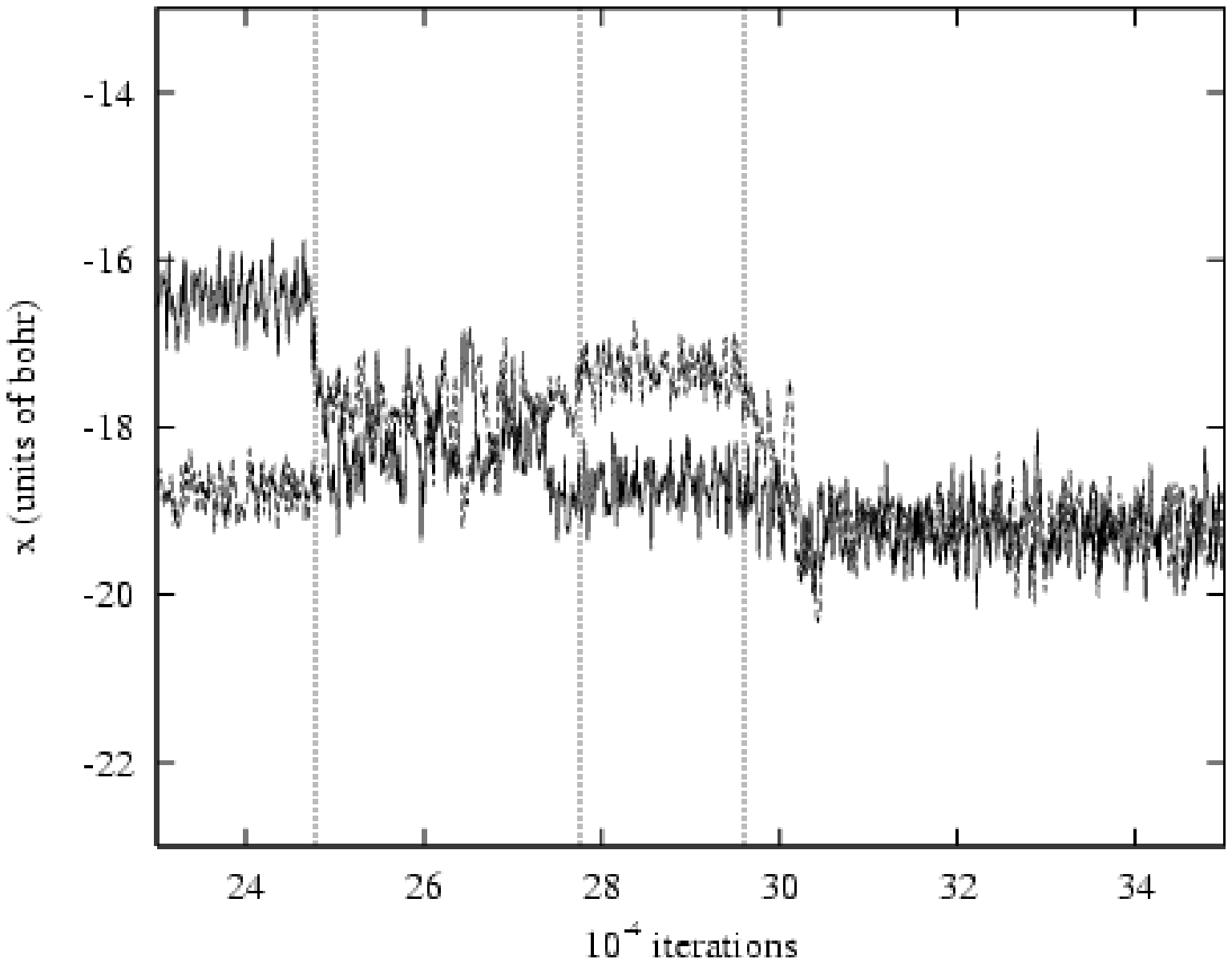}
\caption{The $x$ coordinates of two sodium particles involved in three separate
         transits over a period of $120,000$ iterations.  The transit times are
         the same as in Figs.\ \ref{fig6} and \ref{fig7}.}
\label{fig8}
\end{figure}

Figs.\ \ref{fig6}-\ref{fig8} show the Cartesian coordinates of two 
particles over a common time period.  
The transits in $z$, Fig.\ \ref{fig6}, are sharp and perfectly
correlated in time.  The same holds for the transits in $y$, Fig.\
\ref{fig7}, but both particles show a long postcursor drift following
the third transit.  In the $x$ coordinate, Fig.\ \ref{fig8}, the shift
of the lower particle in the first transit lags the common transit
time by $3\tau$, and this lag is included in our determination of the
average transit width.  Preceding the second transit by a time of
$12\tau$, the lower particle is involved in a transit with a separate
group of other particles.  (Since the transit does not involve both
particles, its time is not plotted in the Figure.)  This transit, at
approximately 275,000 timesteps, produces no discernible shift in the
particle's $y$ and $z$ coordinates.  Then at the second transit, the
upper particle shows a small shift, while the lower particle shows
none.  Of course, both particles move significantly in $y$ and $z$ at
the second transit (Figs.\ \ref{fig6} and \ref{fig7}).  Finally, both
particles show a long drift following the third transit.  We have not
seen such a long postcursor in any other transit.

\section{Discussion}
\label{discuss}

For a binary soft sphere mixture, Miyagawa et al.\ \cite{x} found
correlated jumps in the rms displacement of time-averaged positions of
single particles.  Despite initial appearances, their results are in
fact markedly different from ours.  First, they averaged particle
positions over a time of several vibrational periods, so that motion
on a shorter timescale was not resolved.  Second, they found very
large jumps, around one nearest-neighbor distance, and they found that
several atoms jump at successive times by permuting their positions.
Wahnstr\"{o}m \cite{y} studied a binary Lennard-Jones mixture, and
observed sharp jumps in the magnitude of the displacement of a single
particle as a function of time.  These jumps also do not appear to be
related to the transits we observe.  In Wahnstr\"{o}m's system, the
jumping particle was almost always one of the smaller particles, the
jump distance was at least as large as the nearest-neighbor distance,
and a jumping particle had a tendency to jump back to its original
position after a short time.  It is possible that the nature of the
jumps seen by Miyagawa et al.\ and by Wahnstr\"{o}m is more
characteristic of a dense gas than a liquid.

An observation of cooperative particle motion, via computer
simulation, was reported by Donati et al.\ \cite{19}.  They worked
with a binary mixture of Lennard-Jones particles, at temperatures well
above the glass transition temperature, and observed particle
positions at two different times, separated by a period long compared
to $\tau$.  Between the two observation times, groups of particles
moved a distance on the order of $R_1$ along string-like paths, with
each particle tending to move into the original position of its next
neighbor along the string.  In comparison, in our transits the
particles move a distance noticeably less than $R_1$.  Oligschleger
and Schober \cite{20} studied a system with a soft repulsive potential
at very low temperatures, down to $2.5\%$ of the glass transition
temperature, where they observed jumps in the system rms displacement
versus time.  These jumps corresponded to the motion of particles in
chainlike configurations, where each particle moved only a fraction of
the nearest neighbor distance.  In contrast, our systems do not
exhibit \mbox{transits} at such low temperatures.  Again for a binary
mixture of Lennard-Jones particles, Schr{\o}der et al.\ \cite{104}
used the technique of inherent dynamics to find transitions between
inherent structures that correspond to cooperative string-like
rearrangements of groups of particles moving distances smaller than
the nearest neighbor distance (their Fig.\ 8).  These authors also
found that the distribution of displacements of the equilibrium
positions in such transitions contains a large number of particles
which move a very small distance (their Fig.\ 7).  It is possible that
a similar distribution applies to the monatomic systems studied here.
An important difference between our transits and the motion reported
in \cite{104}, \cite{19}, and \cite{20} is that our transiting groups
do not have string-like configurations, but are more isotropic, albeit
still quite irregular.

In summary, we have observed transits as they appear in the
fluctuating graphs of the particle coordinates in equilibrium
monatomic MD systems.  Each transit is a correlated simultaneous shift
in the equilibrium positions of a small local group of particles.  The
average shift of the equilibrium position of a single particle is
around $0.4\,R_1$ in our Ar system and around $0.25\,R_1$ in our Na
system.  Occasionally a graph of coordinate versus time for a single
particle will show a precursor, or postcursor, extending several
$\tau$ away from the main-group transit time.  The average transit
duration, for a single particle or for the entire group, is roughly
$\tau$ in either Ar or Na, and the precursors and postcursors are
included in this average.

Each precursor or postcursor appears as a segment of the particle
coordinate graph where the mean of the coordinate drifts for a time of
several $\tau$.  In all our calculations, no such drift occurred {\em
except} in connection with a transit.  In our view, the transit itself
is the primary step of diffusive motion.  Nevertheless, the precursors
and postcursors, when they appear, are a part of the equilibrium
diffusive motion, and it would be interesting to further study their
role.

The transits we have observed are isolated events; their duration is
short compared to the time between them.  At higher temperatures the
transits will occur at higher rates.  An important assumption of
liquid dynamics theory is that the motion between random valleys is
accomplished by the same kind of transits observed here, at least to a
first approximation, even though the transit rate in the liquid state
is so high that each particle is involved in a transit approximately once
in every time interval $\tau$.  This hypothesis, as well as other more 
detailed properties of transits, will be investigated in future work.

\end{document}